\documentclass[10pt]{emulateapj}

\usepackage{times,color,natbib,url,amsmath}
\usepackage{graphicx,float,psfrag}
\usepackage{tabularx}
\usepackage{latexsym,amsmath,amssymb}
\usepackage{natbib}
\usepackage{verbatim}
\usepackage{multirow}
\usepackage{courier}

\bibpunct{(}{)}{;}{a}{}{,}

\newcommand {\xmm} {\textsl{XMM-Newton}}
\newcommand {\chandra} {\textsl{Chandra}}

\newcommand {\fermi} {\textsl{Fermi}}
\newcommand {\hess} {H.E.S.S.}

\def \rsun {\ifmmode$R$_{\odot}\else R$_{\odot}$}

\def \hcm {\hbox {\ifmmode $ atoms cm$^{-2}\else atoms cm$^{-2}$\fi}}

\def\approxgt{\mathrel{\hbox{\rlap{\lower.55ex \hbox {$\sim$}}
        \kern-.3em \raise.4ex \hbox{$>$}}}}
\def\approxlt{\mathrel{\hbox{\rlap{\lower.55ex \hbox {$\sim$}}
        \kern-.3em \raise.4ex \hbox{$<$}}}}

\def \arcmin {\hbox{$^\prime$}}
\def \arcsec {\hbox{$^{\prime\prime}$}}

\def \src{Swift~J1834.9$-$0846}

\begin{document}

\title{The wind nebula around magnetar \src}
%
%

\author{G.~Younes$^{1}$, C.~Kouveliotou$^{1}$, O.~Kargaltsev$^{1}$,
  R.~Gill$^{2}$, J.~Granot$^{2}$, A.~L.~Watts$^{3}$, J.~Gelfand$^{4}$,
  M.~G.~Baring$^{5}$, A.~Harding$^{6}$, G.~G.~Pavlov$^7$,
  A.~J.~van~der~Horst$^{1}$, D.~Huppenkothen$^{8,9}$,
  E.~G\"o\u{g}\"u\c{s}$^{10}$, L.~Lin$^{11}$, O.~J.~Roberts$^{12}$}

\affil{
 $^1$ Department of Physics, The George Washington University, Washington, DC 20052, USA \\
 $^2$ Department of Natural Sciences, The Open University of Israel, 1 University Road, P.O. Box 808, Ra\'anana 43537, Israel \\
 $^3$ Astronomical Institute "Anton Pannekoek," University of Amsterdam, Postbus 94249, 1090 GE Amsterdam, The Netherlands \\
 $^4$ NYU Abu Dhabi, P.O. Box 903, New York, NY, 10276, USA \\
 $^5$ Department of Physics and Astronomy, Rice University, MS-108, P.O. Box 1892, Houston, TX 77251, USA \\
 $^6$ Astrophysics Science Division, NASA Goddard Space Flight Center, Greenbelt, MD 20771 \\ 
 $^7$ Department of Astronomy \&\ Astrophysics, Pennsylvania State University, 525 Davey Lab, University Park, PA 16802, USA \\
 $^8$ Center for Data Science, New York University, 726 Broadway, 7th Floor, New York, NY 10003 \\
 $^9$ Center for Cosmology and Particle Physics, Department of Physics, New York University, 4 Washington Place, New York, NY 10003, USA \\
 $^{10}$ Sabanc\i~University, Orhanl\i-Tuzla, \.Istanbul 34956, Turkey \\
 $^{11}$ Department of Astronomy, Beijing Normal University, Beijing
 China 100875 \\
 $^{12}$ School of Physics, University College Dublin, Stillorgan Road, Belfield, Dublin 4, Ireland
}

\begin{abstract}

We report on the analysis of two deep \xmm\ observations of the
magnetar \src\ and its surrounding extended emission taken in March
2014 and October 2014, 2.5 and 3.1 years after the source went into
outburst.  The magnetar is only weakly detected in the first observation
with an absorption corrected flux $F_{\rm
  0.5-10~keV}\approx4\times10^{-14}$~erg~s$^{-1}$~cm$^{-2}$, and a 
$3\sigma$ upper limit during the second observation of about
$3\times10^{-14}$~erg~s$^{-1}$~cm$^{-2}$. This flux level is more 
than 3 orders of magnitude lower than the flux measured at the
outburst onset on September 2011. The extended emission, centered at
the magnetar position and elongated towards the south-west, is clearly
seen in both observations; it is best fit by a highly absorbed
power-law (PL), with a hydrogen column density of $N_{\rm
  H}=8.0\times10^{22}$~cm$^{-2}$ and PL photon index 
$\Gamma=2.2\pm0.2$. Its flux is constant between the two observations
at $F_{\rm 0.5-10~keV}=1.3\times10^{-12}$~erg~s$^{-1}$~cm$^{-2}$. We
find no statistically significant changes in the spectral shape or the flux of
this extended emission over a period of 9 years from 2005 to 2014.
These new results strongly support the extended emission nature
as a wind nebula and firmly establish \src\ as the first magnetar to show
a surrounding wind nebula. Further, our results imply that such
nebulae are no longer exclusive to rotation-powered pulsars and narrow
the gap between these two sub-populations of isolated neutron 
stars. The size and spectrum of the nebula are compatible with those
of pulsar-wind nebulae but its radiative efficiency $\eta_{\rm
  X}=L_{\rm X}/\dot{E}\approx0.1$ is markedly high, possibly pointing
to an additional wind component in \src.

\end{abstract}

\section{Introduction}
\label{Intro}

Magnetars represent a sub-class of isolated neutron stars (NSs) with a
unique set of observational properties. They often have long spin
periods ($P\sim2-12$~s) and large spin-down rates
($\dot{P}\sim10^{-13}-10^{-10}$). They are usually observed as
bright X-ray sources with luminosities, $L_{\rm
  X}\sim10^{32}-10^{36}$~erg~s$^{-1}$, larger than their
corresponding rotational energy losses ($-\dot{E}_{\rm 
  rot}=(2\pi)^2I\dot{P}/P^3\sim10^{30}-10^{35}$~erg~s$^{-1}$, where
$I$ is the NS moment of inertia, $I\approx10^{45}$~g~cm$^{2}$).
Almost all have been observed to emit short ($\sim0.1$~s), bright
($E_{\rm burst}\sim10^{37}-10^{40}$~erg), hard X-ray bursts \citep[see
][for reviews]{mereghetti15:mag,turolla15:mag}. Assuming dipole
braking, the majority of magnetar timing properties indicate strong
surface dipole magnetic fields ($B\gtrsim B_{crit}$, where
$B_{crit}=4.4\times10^{13}$~G is the electron quantum critical
field), while their internal magnetic fields are thought to be even larger
\citep{thompson95MNRAS:GF}. The decay of their internal and external
magnetic fields represent their dominant energy reservoir, powering
their persistent emission as well as their bursting activity \citep{ 
  thompson95MNRAS:GF,thompson96ApJ:magnetar,thompson02ApJ:magnetars,
  beloborodov09ApJ,dallosso12MNRAS}. Finally, a few magnetars have also shown pulsed
radio emission \citep{camilo06Natur:1810, camilo07ApJ:1550,
  torne15MNRAS:1745,rea12ApJ:radiomag}.

In the last decade, several observational results have demonstrated
that the above properties are neither exclusively seen in magnetars
nor solely attributed to super strong surface dipole fields ($B\gtrsim
B_{crit}$). For the purposes of this study, we single out below two of these results. 

PSR~J1846$-$0258 is a $0.3$~s rotation-powered pulsar
(RPP) located inside the supernova remnant (SNR) Kes 75. Its
spin-down rate implies a surface dipole magnetic field
$B=4.9\times10^{13}$~G, on the boundary between RPPs
and classical magnetars \citep{gotthelf00ApJ:psr1846}. Unlike
magnetars, however, PSR~J1846$-$0258 has a large rotational energy
loss rate, $\dot{E}=8.1\times10^{36}$~erg~s$^{-1}$, well above its
persistent X-ray luminosity, $L_{\rm X}=4.1\times10^{34}$~erg~s$^{-1}$. Its 
rotational energy loss also powers a bright pulsar wind nebula (PWN),
$L_{\rm X,~PWN}=1.4\times10^{35}$~erg~s$^{-1}$, with an X-ray efficiency,
$\eta_{\rm PWN}=L_{\rm X}/\dot{E}=2$\%, somewhat high but not unusual
for a young RPP \citep{ng08ApJ:psr1846}. The source spindown age is $\tau =
P/(n-1)\dot{P} = 884\;$yr for a measured breaking index of $n=2.65$
\citep{livingstone06ApJ:kes75}. \citet{gavriil08Sci:kes75}
reported the discovery of short hard X-ray bursts from
PSR~J1846$-$0258, a trademark of typical magnetar sources. The bursts
were accompanied by flux enhancement and timing noise, also typical
properties of magnetars. These observational results demonstrated that
an otherwise typical RPP can in fact show typical magnetar properties,
bridging the  gap between the two NS sub-populations. 

The discovery in 2009 of SGR~J0418+5729 strengthened the above
conclusion. The source was detected after it emitted two 
short hard X-ray bursts \citep{vanderhorst10ApJ:0418}, and exhibited typical
magnetar-like properties: a period of
$\sim9$~s, X-ray flux enhancement soon after the bursts, and a
quasi-exponential flux decay in the following months \citep{
  esposito10MNRAS:0418,rea13ApJ:0418}. A spin-down rate
could only be measured after 3 years of observations and it was
found to be the lowest of any magnetar,
$\dot{P}=4\times10^{-15}$~s~s$^{-1}$, implying a surface dipole field
$B=6\times10^{12}$~G \citep{rea2010Sci:0418}. This field is well within
the range of regular RPPs, indicating that a strong dipole field is
not a requirement for displaying magnetar-like properties in an
isolated NS. 

While the above observational results demonstrate a possible link
between RPPs and magnetars, there exists one RPP property, which has
not thus far been identified in a typical magnetar. Most RPPs possess a large rotational energy loss
rate that powers a relativistic particle wind, often seen
as a PWN, whose X-ray emission is the result of synchrotron radiation
of the shocked wind \citep{kaspi06csxs:PWN,gaensler06ARA:PWN,
  kargaltsev08PWN}. Magnetars, on the other hand, have rotational
energy loss rates on average about 2 orders of magnitude smaller than RPPs
\citep[although with some overlap in their distributions,][]{
  olausen14ApJS:magCat}, making the production of a rotationally
powered nebula less likely. Magnetars, however, are thought to
produce particle outflows, either steady or released during bursting
episodes \citep{thompson98PhRvD:mag,harding96AIPC,harding99ApJ:mag,
  tong13ApJ:wind}, for which the only observational examples is the
transient radio emission detected from SGRs~1900$+$14 and 1806$-$20
following their 1999 and 2004 giant flares, respectively 
\citep{frail99Natur:sgr1900,gaensler05Natur:1806}.

\src\ is a typical magnetar, discovered on 2011 August 7, when it
emitted a short hard X-ray burst. Follow-up X-ray observations
revealed a spin period $P=2.48$~s and a spin-down rate
$\dot{P}=7.96\times10^{-12}$~s~s$^{-1}$, implying a surface
dipole magnetar field strength $B=1.4\times10^{14}$~G (at the equator)
and a rotational energy loss rate, $\dot{E}_{\rm
  rot}=2.1\times10^{34}$~erg~s$^{-1}$. Following the burst, the source
X-ray flux increased by more than 3 orders of magnitude and decayed
quasi-exponentially in the following months
\citep{kargaltsev12apj:1834,esposito13MNRAS:1834}. \xmm\  observations of \src\ in 2011, a month after the source went
into outburst, showed a very unusual extended emission around the
magnetar \citep[][Y+12 hereafter]{younes12ApJ:1834}. This
emission, centered at the source position, was asymmetrical,
extending to the south-west of the magnetar. Moreover, the same
extended emission was detected in an archival \xmm\ observation 6
years earlier at a similar flux level (albeit with large
uncertainties), while the magnetar was $\sim$23 times fainter
(Y+12). Due to the above unusual properties, Y+12 conjectured that this
extended emission might be a wind nebula powered by the magnetar.
 \citet{esposito13MNRAS:1834} later argued that this emission was a
dust scattering halo occurring in a giant molecular cloud located along
the line of sight \citep{tian07ApJ:1834}. They suggested that a
previous outburst from the source prior to 2005 might be responsible
for its earlier detection. The asymmetrical shape was attributed to
non-uniformity in the dust distribution.

In this paper, we report on the analysis of two deep \xmm\
observations of  \src\ and its associated extended emission taken in
March 2014 and in October 2014, 2.5 and 3.1 years after the source
went into outburst. We present the observations and data reduction in
Section~\ref{obs}, and report our analyses results in
Section~\ref{res}. We discuss our findings in Section~\ref{discuss}.
We assume \src\ is at a distance of $d = 4~D_{\rm 4~kpc}$ considering a
likely association with the SNR W41, given its location at the
geometrical center of the remnant \citep{tian07ApJ:1834,leahy08AJ:w41}.

\section{Observations and data reduction}
\label{obs}

We observed \src\ with \xmm\ on two different dates. The first
observation started on 2014 March 16, for a total exposure of 94.9
ks. The second observation took place seven months later, on 2014
October 16, for a total of 85.0 ks. During both observations, the EPIC-PN
\citep{struder01aa} camera operated in extended full-frame mode,
using the medium filter, while the MOS cameras operated in
full-frame mode. We used the Science Analysis System (SAS) version
14.0.0, and HEASOFT version 6.16 for the reduction and analysis of all
data products. Data were selected using event patterns 0$-$4 and 0$-$12
for the PN and MOS cameras, respectively, during only good X-ray events
(``FLAG$=$0''). For both observations, we excluded intervals of
enhanced particle background, only accepting these for which the count
rate above $10$~keV for the entire PN and MOS fields of view did not
exceed 0.4 and 0.35 counts~s$^{-1}$,
respectively\footnote{http://xmm.esac.esa.int/sas/current/documentation/threads/EPIC\_filterbackground.shtml}. Table~\ref{logObs}
lists the log of the 2 \xmm\ observations.

To perform our spectral analysis, source events were extracted from specific regions as
described in Section~\ref{specana} and shown in
Figure~\ref{firstLookIm}. Background events were extracted from a
5\arcmin-radius circle on the same CCD as the source, excluding point
sources as derived from a source detection algorithm
(Section~\ref{imana}). The task {\sl backscale} was used to calculate
the exact area of the source and background regions, correcting for
excluded point sources, CCD gaps, and bad pixels. We generated
response matrix files using the SAS task {\sl rmfgen}, while ancillary
response files were generated using the SAS task {\sl arfgen}. All
spectra were created in the energy range 0.5--10~keV.

XSPEC \citep{arnaud96conf} version 12.8.2 was used for our
analysis. The photo-electric cross-sections of \citet{
  verner96ApJ:crossSect} and the abundances of \citet{wilms00ApJ} were
used to account for absorption by neutral gas. For all spectral fits
using all three cameras of the EPIC detector, we added a
multiplicative constant normalization, frozen to 1 for the PN and
allowed to vary for MOS1 and MOS2, to take into account any
calibration uncertainties between the three instruments. We found a
3-5\% variation in the MOS1 and MOS2 normalizations relative to
PN. All quoted uncertainties in this study are at the $1\sigma$ level,
unless otherwise noted.

\begin{table}[]
\caption{Log of the \xmm\ observations}
\label{logObs}
\newcommand\T{\rule{0pt}{2.6ex}}
\newcommand\B{\rule[-1.2ex]{0pt}{0pt}}
\begin{center}{
\resizebox{0.48\textwidth}{!}{
\begin{tabular}{l c c c}
\hline
\hline
Observation ID   \T\B & Date & Instrument & Good Time Intervals \\
                           \T\B &         &                    &      (ks)     \\
\hline
0723270101 \T & 2014-03-16 & PN & 70.0 \\
                      \T & 2014-03-16 & MOS1 & 87.7 \\
                      \T & 2014-03-16 & MOS2 & 88.5 \\
0743020201 \T & 2014-10-16 & PN & 60.0 \\
                      \T & 2014-10-16 & MOS1 & 78.0 \\
                      \T & 2014-10-16 & MOS2 & 77.7 \\
\hline
\end{tabular}}}
\end{center}
\end{table}

\section{Results}
\label{res}

\subsection{Imaging analysis}
\label{imana}

\begin{figure*}[]
\begin{center}
\includegraphics[angle=0,width=0.98\textwidth]{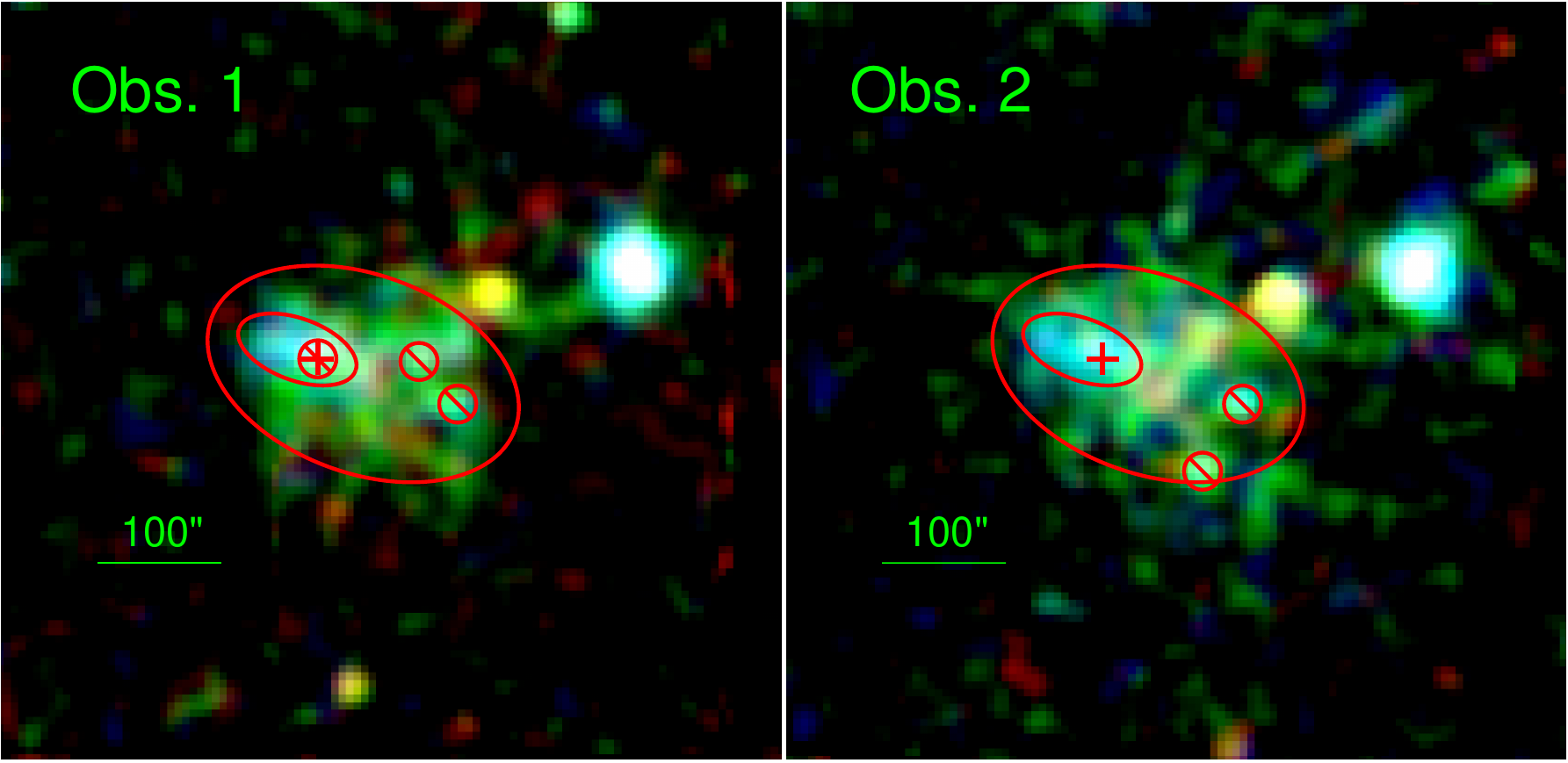}\\
\includegraphics[angle=0,width=0.98\textwidth]{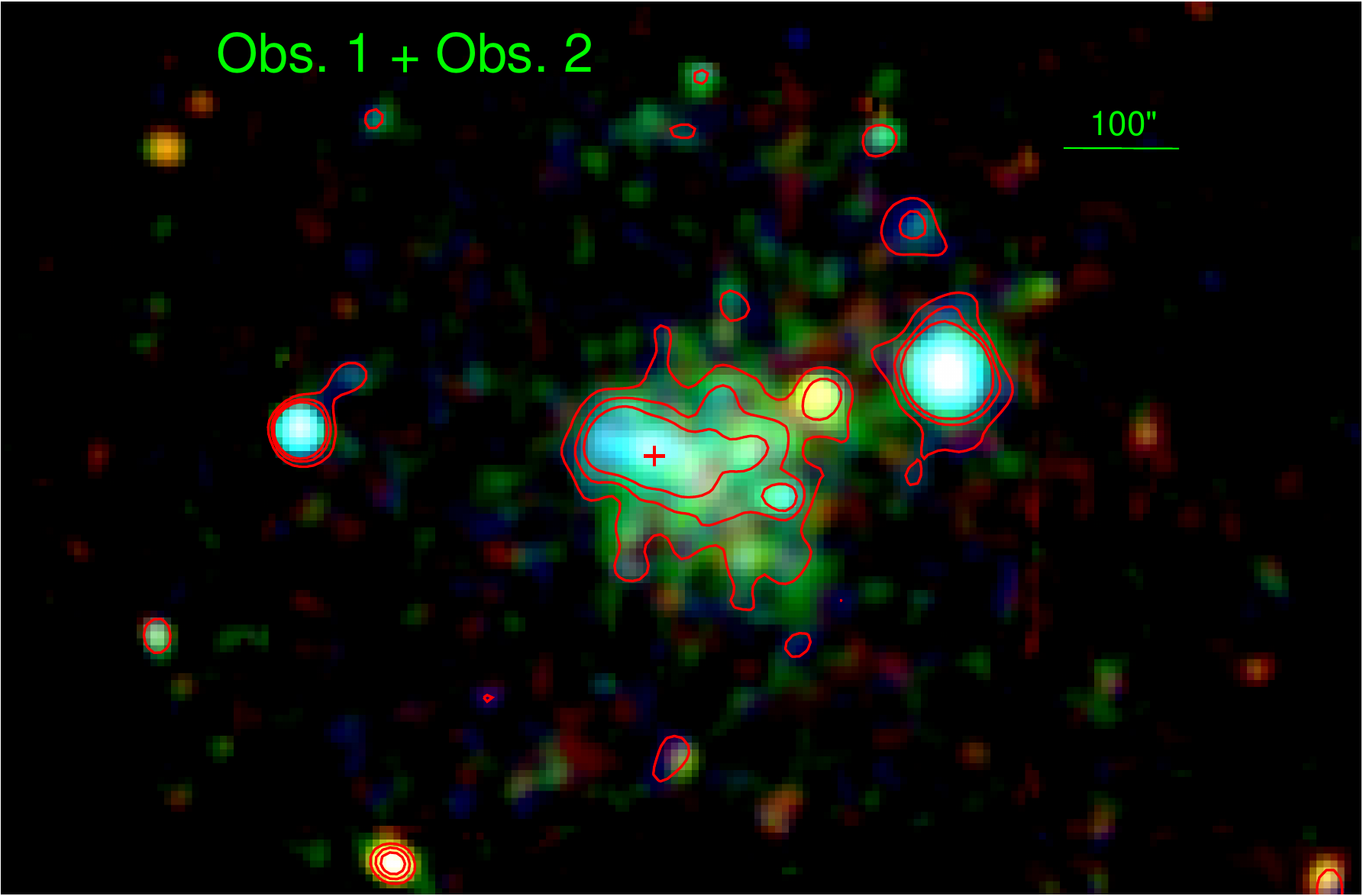}
\caption{{\sl Upper panels.} PN+MOS1+MOS2 exposure-map corrected RGB images ($2-3$~keV in red, $3-4.5$~keV in green, $4.5-10$~keV in blue) of the extended emission around \src\ during obs.~1 (2014 March 16,
  left) and obs.~2 (2014 October 16, right). The red
  cross indicates the magnetar position, which is weakly detected in
  obs.~1 above the strong extended emission. The inner
  ($25$\arcsec$\times50$\arcsec\ minor and major axes) and outer
  ($80$\arcsec$\times130$\arcsec\ minor and major axes) ellipses are
  regions used to investigate spatial-spectral evolution within
  the nebula. Red crossed-out circles are weak point sources within
  the extended emission excluded from any imaging and spectral
  analyses. {\sl Lower panel.} PN+MOS1+MOS2 combined images from the
  two observations. The contours are at the 2.5, 3.0, and 3.5 $\sigma$
  levels. The red cross indicates the magnetar position. See text for
  more details.} 
\label{firstLookIm}
\end{center}
\end{figure*}

We used the {\sl images}
script\footnote{http://xmm.esac.esa.int/external/xmm\_science/gallery/utils/images.shtml}
to produce a cleaned image of the nebula for both \xmm\
observations. This script uses raw event data files, and filters for
high background intervals, removes bad pixels and columns, corrects
for several camera inefficiencies through an exposure map and, finally,
merges PN and MOS data. The script allows the production of these
images in different energy bands, using a specified pixel binning and
smoothing radius. Figure~\ref{firstLookIm} shows the results of the
script for observation 0723270101 (obs.~1 hereinafter, upper-left
panel) and 0743020201 (obs.~2 hereinafter, upper-right panel). We use
three energy bands to produce these images, 2-3 (red), 3-4.5 (green),
and 4.5-10~keV (blue). The 2~keV lower limit was set because the source is highly absorbed, and hence, data below 2~keV are
mostly due to foreground noise. The images are also binned to
6\arcsec/pixel and smoothed with a FWHM of 20\arcsec. The extended
emission, clearly present in both images around the magnetar
position (marked as a red cross), is remarkably similar in the
two observations. We show in the lower panel of
Figure~\ref{firstLookIm} the merged images of the two observations of
the extended emission with the 2.5, 3, and 3.5
$\sigma$ contours. This is the deepest image of the
extended emission around the magnetar \src.

Figure~\ref{firstLookIm} shows a clear trend in hardness,
with the inner part of the extended emission appearing harder than its
outskirts. To identify and remove the contribution of any background (or foreground) point sources
within the field of the extended emission, we ran two source detection
algorithms. The
\texttt{edetectchain}\footnote{http://xmm.esac.esa.int/sas/current/doc/edetect\_chain/edetect\_chain.html}
uses an exposure corrected image to look for sources within a 5$\times$5
pixel box using a surrounding background of a $2$ pixel box beyond the
source in all input images (e.g., PN, MOS1, and MOS2) simultaneously.
It then masks the sources found, and creates a background map of the
field-of-view through a 2-D spline fit. Finally, using the
values from this background map, it searches for sources within a
5$\times$5 pixel box in all input images simultaneously. The
\texttt{edetectchain} algorithm also performs point source or
extended emission fits to each of the sources found through the task
\texttt{emldetect}. The other detection algorithm we ran on the data
(PN, MOS1, and MOS2 separately) is
\texttt{wavdetect}\footnote{http://cxc.harvard.edu/ciao/threads/wavdetect/}. This
algorithm correlates a ``Ricker wavelet'' (``Mexican Hat wavelet'')
function to a given 2-D image. Pixels with large positive correlation
and a low significance ($<2\times10^{-6}$) are flagged as sources and removed from the data, and the
same correlations are performed again until no more sources are found.
This algorithm is developed for \chandra\ observations, but it can
also be used with \xmm\ data using the relevant exposure and point spread
function (PSF) map files.

\begin{figure*}[ht]
\begin{center}
\includegraphics[angle=0,width=0.95\textwidth]{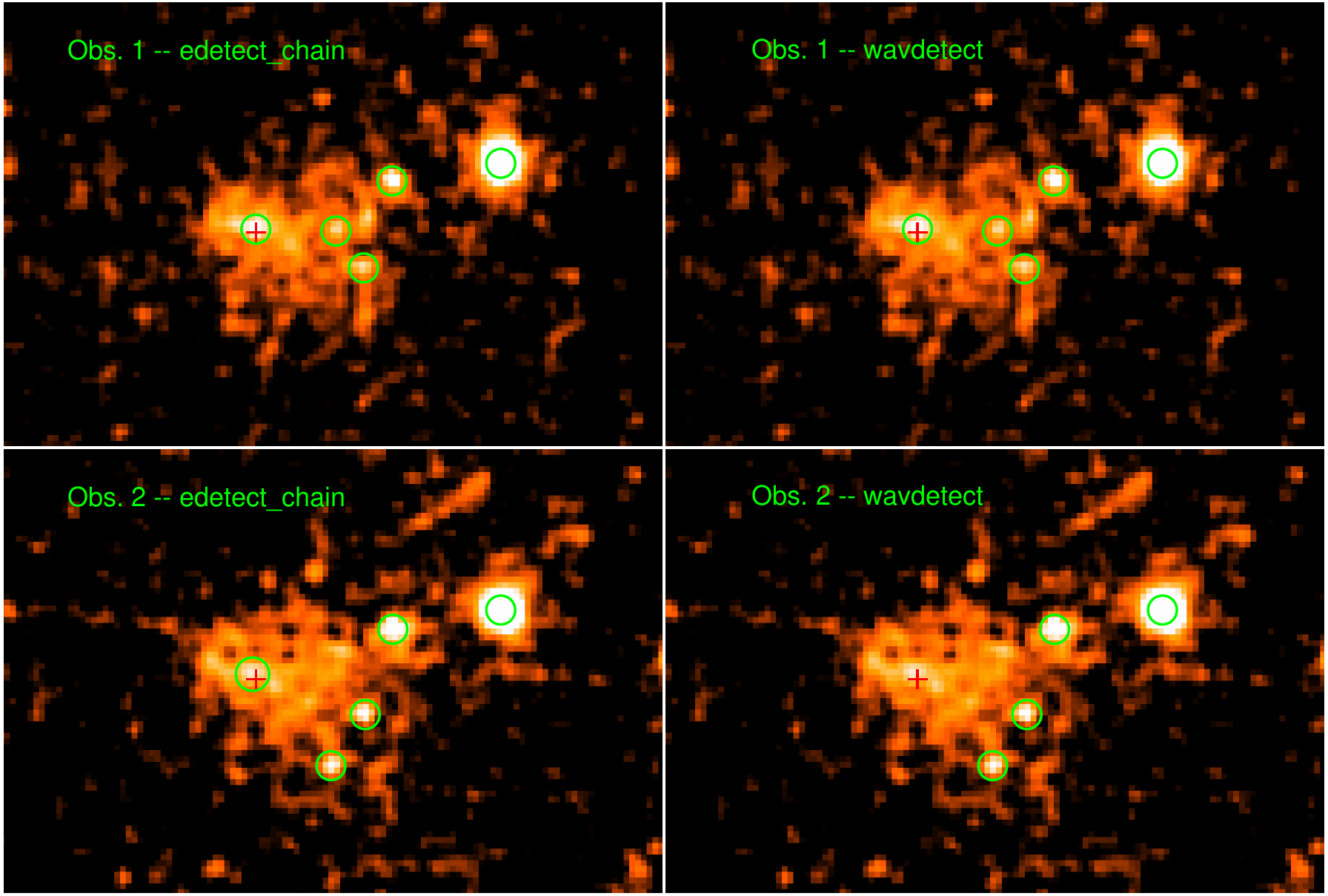}
\caption{\xmm\ EPIC images in the 2-10 keV band, Gaussian
    smoothed with a FWHM of 12\arcsec. {\sl Upper panels.} Obs.~1
    image with the results from the source detection algorithm
    \texttt{edetectchain} (left) and \texttt{wavdetect} 
  (right) overlaid. {\sl Lower panels.} Obs.~2 image with the results
  from the source detection algorithm \texttt{edetectchain} (left)
  and \texttt{wavdetect} (right) overlaid. The position of \src\ is
  marked as a red cross in all panels. The difference in the point
  sources detected between the two observations is likely the result
  of weak X-ray transients.}
\label{rawima}
\end{center}
\end{figure*}

Figure~\ref{rawima} shows the PN, MOS1, and MOS2 exposure corrected
combined images for obs.~1 (upper panels) and obs.~2 (lower panels) in
the 2-10~keV range, along with the results from both source-detection
algorithms. Both algorithms produce similar results, except that the
\texttt{edetectchain} algorithm flags a source at the position of the
magnetar in obs.~2, while the wavdetect algorithm does not. We note
that the source detected using \texttt{edetectchain} at the magnetar
position in both observations is flagged as extended. The faint
  X-ray sources detected in the two observations are likely weak
  X-ray transients \citep[e.g.,][]{asai98PASJ:XTrans,campana98AARv:XTrans}, 
  resulting in different source detections between obs.~1 and obs.~2.

\begin{figure*}[!t]
\begin{center}
\includegraphics[angle=0,width=0.98\textwidth]{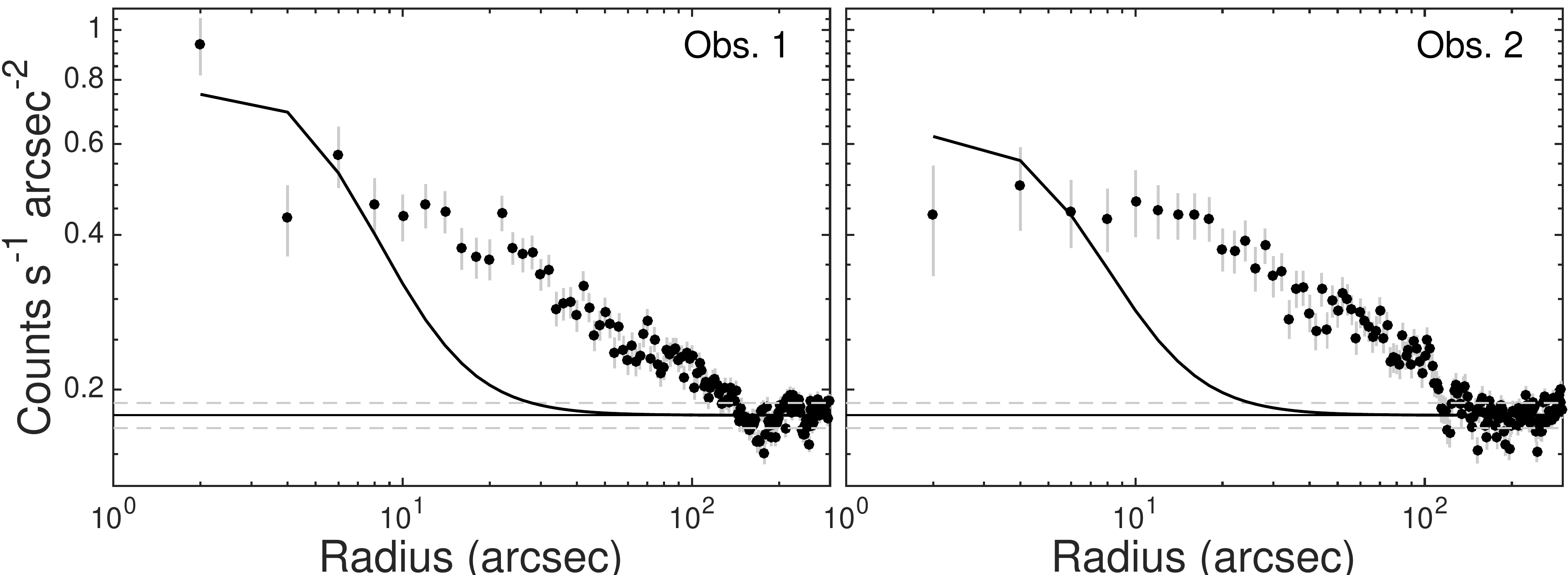}
\caption{The 2-10 keV radial profiles from the \xmm\ obs.~1 ({\sl left
    panel}) and obs.~2 ({\sl right panel}) centered on the position of
  \src. The profile of obs.~1 is normalized to have the same
  background level as obs.~2 (shown as horizontal black solid line with its
  $1\sigma$ deviation as grey dashed lines). The average PSF from all
  three EPIC instruments for a point source at the magnetar position
  is also shown. See text for details.}
\label{radProf}
\end{center}
\end{figure*}

To determine whether the magnetar emission is detected above the
extended-emission level, we estimated the number of
counts and computed the hardness ratios (HRs) in both observations. Using the PN
camera in the 2-10 keV, we estimate the background-corrected number of
counts  in a circle with radius 20\arcsec\ (75\% encircled energy)
around the magnetar position \citep{kargaltsev12apj:1834}. Normalizing
by the Good Time Intervals in both observations and correcting for bad
pixels and CCD gaps, we find $245\pm9$ counts and $196 \pm11$ counts
for obs.~1 and obs.~2, respectively. The difference of $49\pm14$
counts between the two observations represents a $3.5\sigma$
significance. Moreover, the radial profiles of the two observations in
the 2-10 keV range  (Figure~\ref{radProf}) centered at the magnetar
position \citep{kargaltsev12apj:1834} reveal a central excess emission
in obs.~1 compared to obs.~2. Beyond a few arcseconds, the radial
profiles from both observations are similar. We conclude that there is
a $3.5\sigma$ count excess around the magnetar position in obs.~1.

We estimate the HR (HB$_{\rm 4.5-10~keV}$/SB$_{\rm 2-4.5~ keV}$)
within the 20\arcsec\ circle by first
estimating the photon flux in the two energy bands for all instruments
separately. The photon flux is estimated by correcting the number of
counts observed within the 20\arcsec\ circle for CCD gaps and bad
pixels, and then normalizing by the ``LIVETIME'' exposure and the
average detector effective area at the source location in the given
energy band. Finally, we subtracted the background contribution from
these photon fluxes using the background region as described in
Section~\ref{obs}. The photon fluxes in the different energy bands and
the hardness ratios for obs.~1  and obs.~2, are shown in
Table~\ref{photFluxHR}.

We also derived the hardness ratios of two other regions within the
extended emission following the same method as above. The first
region, hereafter the inner-ellipse, is indicated by the smaller red
ellipse in Figure~\ref{firstLookIm}, and the second region, hereafter
the outer-ellipse, is indicated by the larger red ellipse. These
elliptical regions were defined according to their spectral appearance 
in the image. Point sources within each of these two regions (defined
as crossed-out circles with 15\arcsec radii), as derived from
the source detection algorithms, are excluded. The magnetar (shown as
a red circle with a 20\arcsec-radius) was excluded from the
inner-ellipse, and the inner-ellipse was excluded from the
outer-ellipse. These photon fluxes and hardness ratios for obs.~1  and obs.~2, are shown in
Table~\ref{photFluxHR}.

\begin{table}[]
\caption{Photon fluxes and hardness ratios for different locations within
  the extended emission}
\label{photFluxHR}
\newcommand\T{\rule{0pt}{2.6ex}}
\newcommand\B{\rule[-1.2ex]{0pt}{0pt}}
\begin{center}{
\resizebox{0.48\textwidth}{!}{
\begin{tabular}{l c c c}
\hline
\hline
 \T\B & Ph. flux (2-4.5 keV) & Photon flux (4.5-10 keV) & HR \\
 \T\B &  $10^{-6}$~photons~cm$^{-2}$~s$^{-1}$       & $10^{-6}$~photons~cm$^{-2}$~s$^{-1}$ &  \\
\hline
magnetar position (obs.~1) \T & $3.0\pm0.2$  & $5.1\pm0.5$  & $1.7\pm0.2$ \\
magnetar position (obs.~2) \T\B & $2.3\pm0.2$  & $5.5\pm0.6$  & $2.4\pm0.3$ \\
\hline
Inner-ellipse (obs.~1) \T & $2.0\pm0.2$  & $4.7\pm0.5$  & $2.4\pm0.4$ \\
Inner-ellipse (obs.~2) \T & $1.6\pm0.2$  & $4.3\pm0.5$  & $2.7\pm0.5$ \\
Inner-ellipse (average)\T\B & $1.8\pm0.1$  & $4.5\pm0.4$  & $2.5\pm0.3$ \\
\hline
Outer-ellipse (obs.~1) \T & $15.0\pm1.0$  & $13.4\pm1.5$  & $0.9\pm0.1$ \\
Outer-ellipse (obs.~2) \T & $13.3\pm0.7$  & $10.9\pm1.4$  & $0.8\pm0.1$ \\
Outer-ellipse (average)\T\B& $14.2\pm0.6$  & $12.2\pm1.0$  & $0.9\pm0.1$ \\
\hline
\hline
\end{tabular}}}
\end{center}
\end{table}

Two interesting conclusions can be drawn from the results in
Table~\ref{photFluxHR}. First, the HR in a 20\arcsec circle around the
magnetar position indicates a softer spectrum in obs.~1 than in
obs.~2. Moreover, the HR around the magnetar position in obs.~2 is
almost identical to the HR in the inner-ellipse of the two
observations. Hence, given the central excess counts in obs.~1
compared to obs.~2, and the softer spectrum, we conclude that the
magnetar is weakly detected in obs.~1, while it has faded as indicated by the
harder extended emission during obs.~2. Second, the HR of the
outer-ellipse is noticeably smaller than the HR of the inner-ellipse,
indicating spectral softening with increasing distance from the
magnetar position.

\subsection{Spectral analysis}
\label{specana}

We extracted the source spectra from the
area within the big ellipse ($80\arcsec\times130\arcsec$) in
Figure~\ref{firstLookIm} (upper panels), excluding the point
sources detected in each observation. The
background spectrum is extracted from a region as defined in
Section~\ref{obs}. We then subtracted the background spectrum
from the source emission spectrum, as is usually done for point
sources. This approach should be valid for our case since the extent
of the emission is too small to cause any strong vignetting effects. We
used the Cash statistic (C-stat in XSPEC) for our parameter
estimation, and grouped the spectra to have 5 counts per bin. We fit
the PN, MOS1, and MOS2 spectra of obs.~1 and obs.~2 simultaneously
with an absorbed PL model.  We linked all parameters between the
two observations assuming no variability in the extended emission.

We find a PL photon index $\Gamma=2.2\pm0.2$ and an absorbing hydrogen
column density $N_{\rm H}=8.0_{-0.8}^{+0.9}\times10^{22}$~cm$^{-2}$.
Since we are using C-stat for parameter estimation, we used the XSPEC
command
\texttt{goodness}\footnote{https://heasarc.gsfc.nasa.gov/docs/xanadu/xspec/manual/XSgoodness.html} 
to evaluate the goodness of fit. The \texttt{goodness} command
simulates a user-defined number of spectra based on a Gaussian
distribution of the best fit model parameters. It derives the
percentage of simulations with fit statistic lower than that for the
data. In the case where the data are drawn from the model, this
percentage should be around 50\%. Simulating 10000 realizations of our
data based on the above best fit model, we find that 57\% of the
simulated spectra have a fit statistic lower than the best fit
statistic, C-stat=4210.82 for 4140 degrees of freedom (d.o.f.),
implying that our simple model provides a good fit to the data.

\begin{figure}[]
\begin{center}
\includegraphics[angle=0,width=0.49\textwidth]{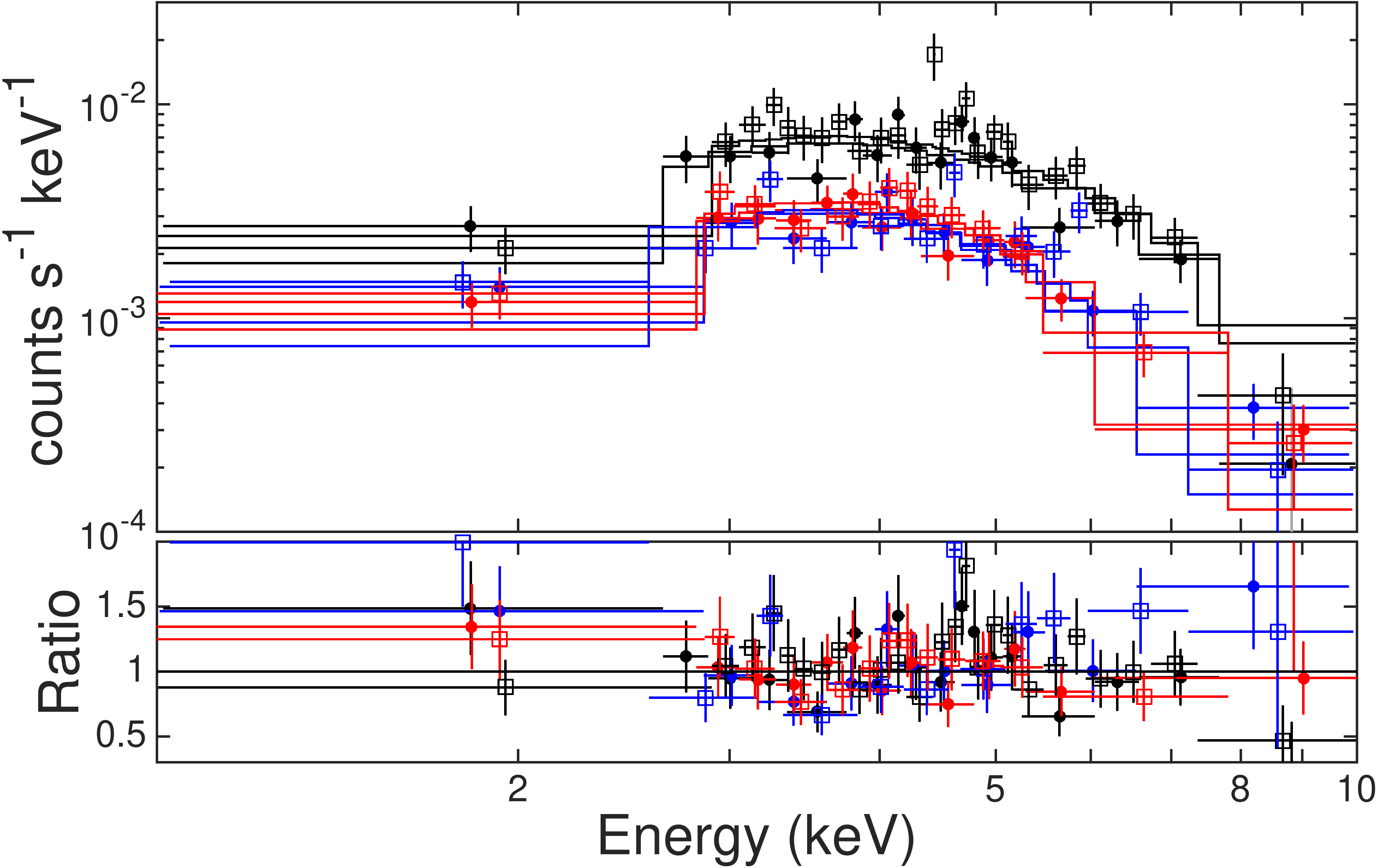}
\caption{{\sl Upper panel.} Data and best fit model of the 2014
    \xmm\ observations of the extended emission around \src. The dots
    and open squares represent obs.~1 and obs.~2, respectively. Black,
    blue, and red are for PN, MOS1, and MOS2, respectively. {\sl Lower
      panel} Data to model ratio. The best fit model in this plot is
    obtained using C-stat; data are rebinned for clarity. See text for
  more details.}
\label{specFit}
\end{center}
\end{figure}

\begin{figure*}[!th]
\begin{center}
\includegraphics[angle=0,width=0.33\textwidth]{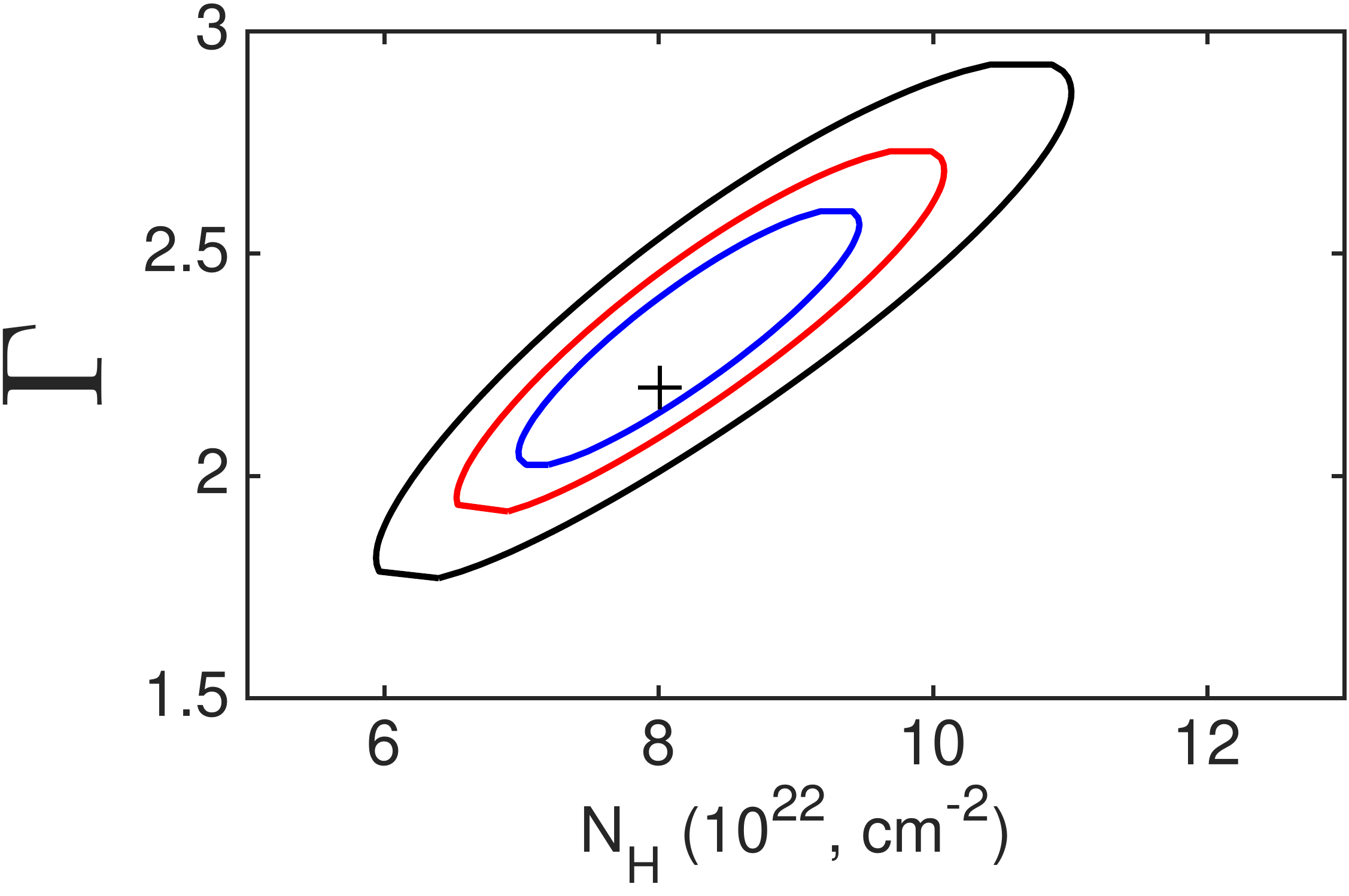}
\includegraphics[angle=0,width=0.33\textwidth]{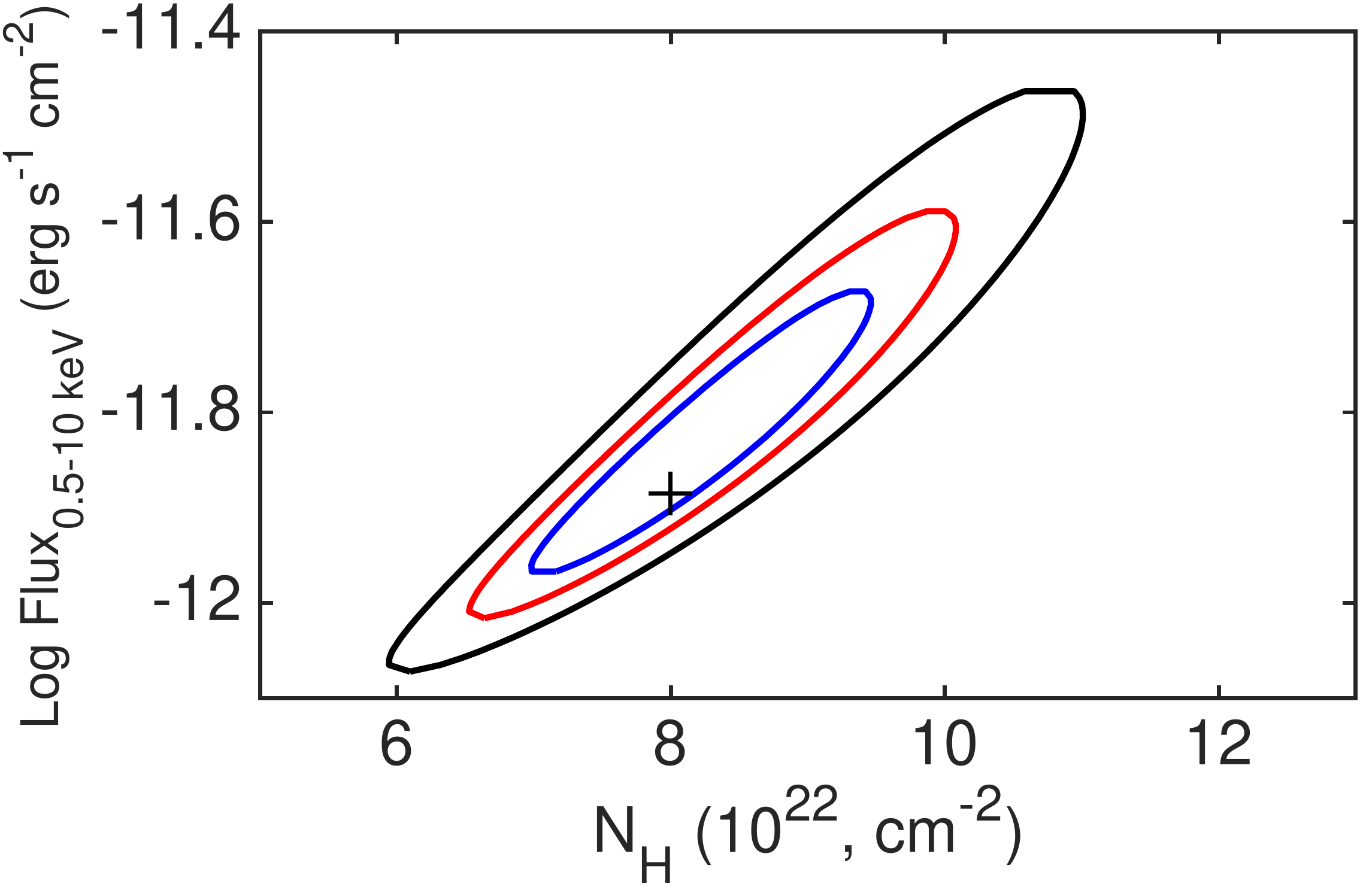}
\includegraphics[angle=90,width=0.33\textwidth]{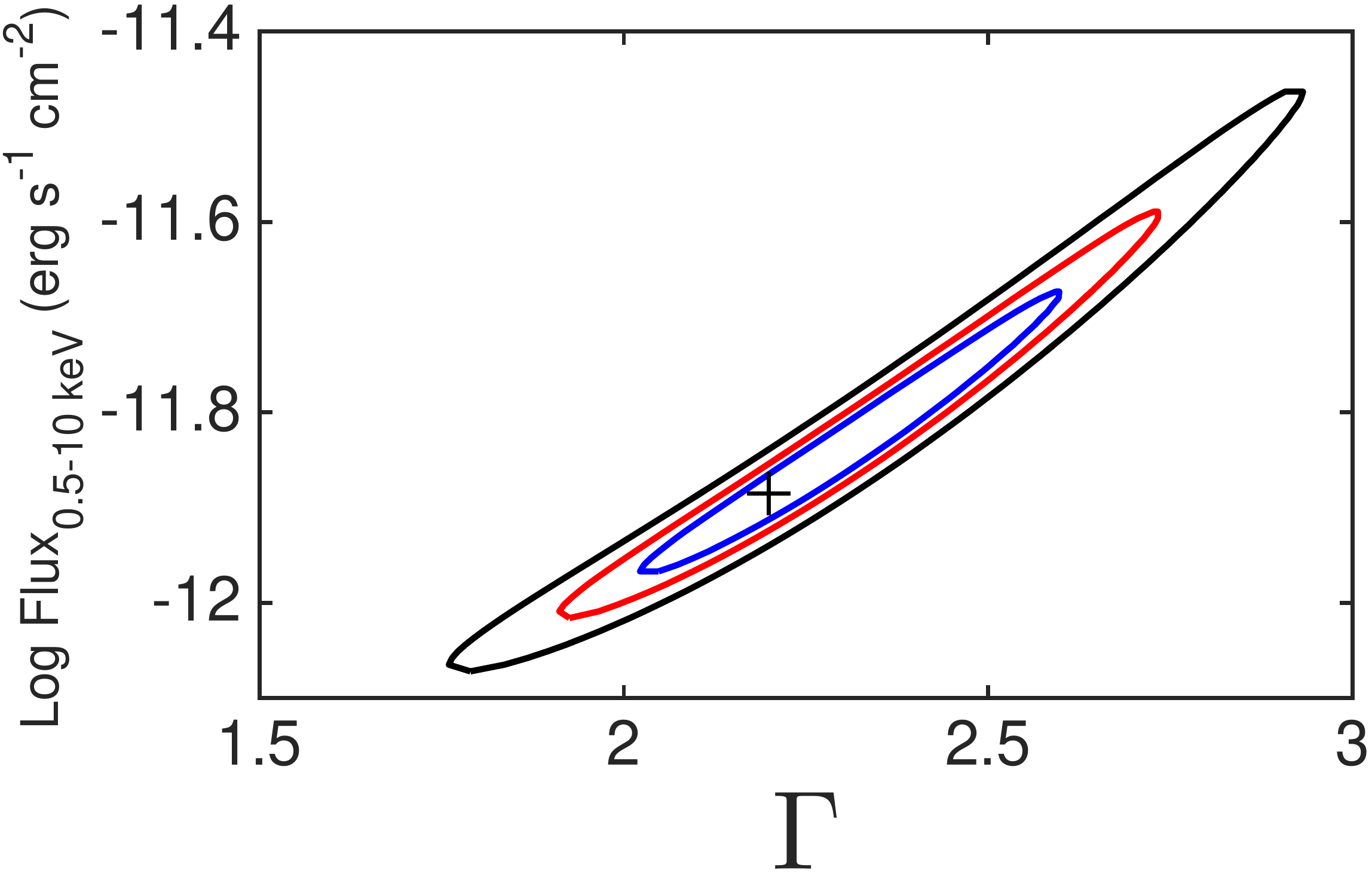}
\caption{{\sl Left panel.} $N_{\rm H}$-$\Gamma$ contours. {\sl Middle
    panel.} $N_{\rm H}$-$\log F_{\rm 0.5-10~keV}$ contours. {\sl Right
    panel.} $\Gamma$-$\log F_{\rm 0.5-10~keV}$ contours. In all three
  panels, the black, red, and blue contours are at the 1$\sigma$,
  2$\sigma$, and 3$\sigma$ levels.}
\label{contours}
\end{center}
\end{figure*}

We find an absorption corrected 0.5-10~keV flux
$F_{\rm 0.5-10~keV}=1.3_{-0.2}^{+0.4}\times10^{-12}$~erg~s$^{-1}$~cm$^{-2}$, which
translates into a luminosity 
$L_{\rm 0.5-10~keV}=2.5_{-0.6}^{+0.7}\times10^{33}$~$D_{\rm
  4~kpc}^2$~erg~s$^{-1}$, assuming a distance to the source $d =
4~D_{\rm 4~kpc}$~kpc.  The spectra and best fit model are shown in
Figure~\ref{specFit}. The 2-D contour plots between $N_{\rm H}$,
$\Gamma$, and $F_{\rm 0.5-10~keV}$ are shown in
Figure~\ref{contours}. The spectral fit results are summarized in
Table~\ref{specParam}.

\begin{table}[!t]
\caption{Nebula PL spectral parameters}
\label{specParam}
\newcommand\T{\rule{0pt}{2.6ex}}
\newcommand\B{\rule[-1.2ex]{0pt}{0pt}}
\begin{center}{
\resizebox{0.49\textwidth}{!}{
\begin{tabular}{c c c c c}
\hline
\hline
Statistic used      \T\B & $N_{\rm H}$  & $\Gamma$ & $F_{\rm 0.5-10~keV}$ & $L_{\rm 0.5-10~keV}^a$ \\
                          \T\B & ($10^{22}$ cm$^{-2}$) & & ($10^{-12}$ erg s$^{-1}$ cm$^{-2}$) & ($10^{33}$ erg s$^{-1}$)\\
\hline
C-stat$^b$      \T\B & $8.0_{-0.8}^{+0.9}$ & $2.2\pm0.2$ & $1.3_{-0.2}^{+0.4}$ & $2.5_{-0.6}^{+0.7}$\\
C-stat$^c$      \T\B & $7.3\pm1.0$ & $2.1\pm0.2$ & $1.1_{-0.3}^{+0.5}$ & $2.1_{-0.6}^{+1.0}$\\
$\chi^2$ \T\B &  $8.0\pm1.0$  &  $2.1\pm0.3$     & $1.2_{-0.2}^{+0.5}$ & $2.3_{-0.5}^{+0.9}$\\
\hline
\end{tabular}}}
\end{center}
\begin{list}{}{}
\item[{\bf Notes.}]$^a$Derived by adopting a 4~kpc distance. $^b$
  Background subtraction method. $^c$ Modelled
  background method.
\end{list}
\end{table}

We also looked for flux variability between obs.~1 and obs.~2 by
leaving the normalization of the PL free to vary between the two
spectra. We find a C-stat of 4212.95 for 4139 d.o.f, and
the normalization of the two spectra are consistent with each other at
the $1\sigma$ level. To establish whether the change in C-stat is
statistically significant, we estimated the Bayesian Information
Criterion (BIC) in both cases. The BIC for the case of linking the
normalization of the PL, i.e., constant flux, is 4260, while the BIC
for a varying PL normalization is 4255. This gives $\Delta BIC=5$,
which implies that the case of the free PL normalization, i.e., a
varying flux, is not statistically preferable over the simpler case,
i.e., constant flux.

Following Y+12, we also performed a spectral analysis by first
modeling the background spectrum, and then including its contribution
to the source spectral model. Hence, we fit the background spectrum
with a combination of two thermal components (for the local hot bubble
and interstellar/intergalactic medium), and 2 PLs, one with a
photon index fixed to 1.5 (assuming unresolved background active
galactic nuclei, e.g., distant quasars and/or nearby low luminosity
active galactic nuclei; \citealt{porquet04aa:pgquasar,sazonov08AA:xrb,
  younes11AA:liner1sXray}) and absorbed by a column density equal to
the average value of the Galactic absorption towards the direction of
the background region, $N_{\rm H}\approx2.0\times10^{22}$~cm$^{2}$. The
temperatures of the two thermal components are $\sim$0.2~keV
and $\sim$1.1~keV; both reasonable for the thermal emission in the
diffuse X-ray background \citep{snowden04ApJ:xmmbackgr,
  snowden08AA:xmmbackgr}. We find $\Gamma\approx0.6$ for the
unabsorbed foreground PL component, which could represent some low
level solar flaring background below our exclusion threshold
(Section~\ref{obs}). We also added Gaussian emission lines to model
the instrumental lines seen in PN and MOS (see the Extended Science
Analysis Software,
ESAS\footnote{http://xmm.esac.esa.int/sas/current/doc/esas/index.html}). After  
establishing the best fit model to the background spectrum, we
included the source spectra from the two observations and added an
absorbed PL component to the total spectral model. We linked the
absorption column density of the background model to that of the
source. The spectral fit results for this absorbed PL component, which
represents the extended emission spectral model, are summarized in
Table~\ref{specParam}. These results are in very good agreement with
the results from our initial method.

Finally, we also performed the spectral fitting using the more
commonly used $\chi^2$ statistics. We binned all spectra to have a
signal to noise ratio of 3, and fit them with an absorbed PL, linking
all parameters together. The fit is remarkably good with $\chi^2=153$
for 160 d.o.f. We find a PL photon index $\Gamma=2.1\pm0.3$ and an
absorbing hydrogen column density $N_{\rm
  H}=(8.0\pm1.0)\times10^{22}$~cm$^{-2}$. These results are in
agreement with the above two methods at the $1\sigma$ confidence
level. The fit parameters along with their uncertainties are
summarized in Table~\ref{specParam}. Using the $\chi^2$ statistics, we
also studied the case of a varying flux between the 2
observations. Letting the PL normalization free to vary, we find a
$\chi^2=149$ for 159 d.o.f. This results in an F-test statistic of 3.2
and a false-rejection probability of 8\%, implying that a variable
flux is not statistically favored over a constant flux.

The imaging analysis of the nebula (Figure~\ref{firstLookIm}) revealed
a softening trend with distance from the central magnetar. To
investigate this trend, we extracted the PN, MOS1, and MOS2 spectra
of the two regions within the nebula as identified in
Section~\ref{imana}, i.e., the inner- and outer-ellipses. We used
C-stat in the fitting process and grouped the spectra to have 5
counts per bin. We fit the spectra simultaneously with an absorbed
PL model, letting the normalization of the PL vary freely. As a first
attempt to model the softening trend, we linked the absorption column
density between the inner and the outer ellipses and let the PL index
free. This assumes that the whole extent of the nebula is equally
absorbed and the softening is due to a change in the spectral
curvature of the photon spectrum. This assumption leads to a good fit
with C-stat of 2859.52 for 2749 d.o.f. We find a common hydrogen
column density $N_{\rm H}=(12\pm2)\times10^{22}$~cm$^{-2}$. The photon
indices of the inner and outer ellipses are $\Gamma_{\rm
  Inn}=1.3\pm0.3$ and $\Gamma_{\rm Out}=2.5\pm0.2$, respectively. We
also tried linking the PL photon index between the inner and outer
ellipses while leaving the hydrogen column density free to vary,
effectively assuming that the softening is due to different absorbing column
towards different parts of the nebula. We find an equally good fit
with C-stat of 2862.06 for 2749 d.o.f.
We find a common PL index $\Gamma=2.3\pm0.2$, while the absorption is
$N_{\rm H, Inn}=(22\pm3)\times10^{22}$~cm$^{-2}$ and $N_{\rm H,
  Out}=(11\pm2)\times10^{22}$~cm$^{-2}$ for the inner and outer
ellipses, respectively. Leaving both the absorption and the photon
index free to vary does not provide any additional improvement to the
fits with C-stat of 2858.97 for 2748 d.o.f. These results are
discussed in Section~\ref{discuss}.

We checked whether an optically-thin thermal component can explain the
extended emission spectral properties (optically-thick thermal
emission is unlikely given the large size of the emission region).
Using the $\chi^2$ statistics, we fit all spectra of the whole nebula
to a hot diffuse gas model (\texttt{APEC} in XSPEC). Fixing the
abundance to solar, we find a statistically acceptable fit with
$\chi^2$ of 163 for 159 d.o.f. We find a very high gas temperature
with a $3\sigma$ lower limit $kT\gtrsim32$~keV. Allowing the
abundance to vary, we find an equally good fit with reasonable gas
temperature $kT=7_{-1}^{+3}$~keV. The abundance, however, is very low
with a $3\sigma$ upper limit of $<0.1$ (in solar units). There
are other local minima that could be found in the $\chi^2$ space
resulting in reasonable gas temperatures ($kT\sim1$~keV) and
abundances (close to solar). These fits, however, are statistically
unacceptable with reduced $\chi^2$ in the range of $1.5-1.6$ for 158
d.o.f.

\begin{figure}[]
\begin{center}
\includegraphics[angle=90,width=0.48\textwidth]{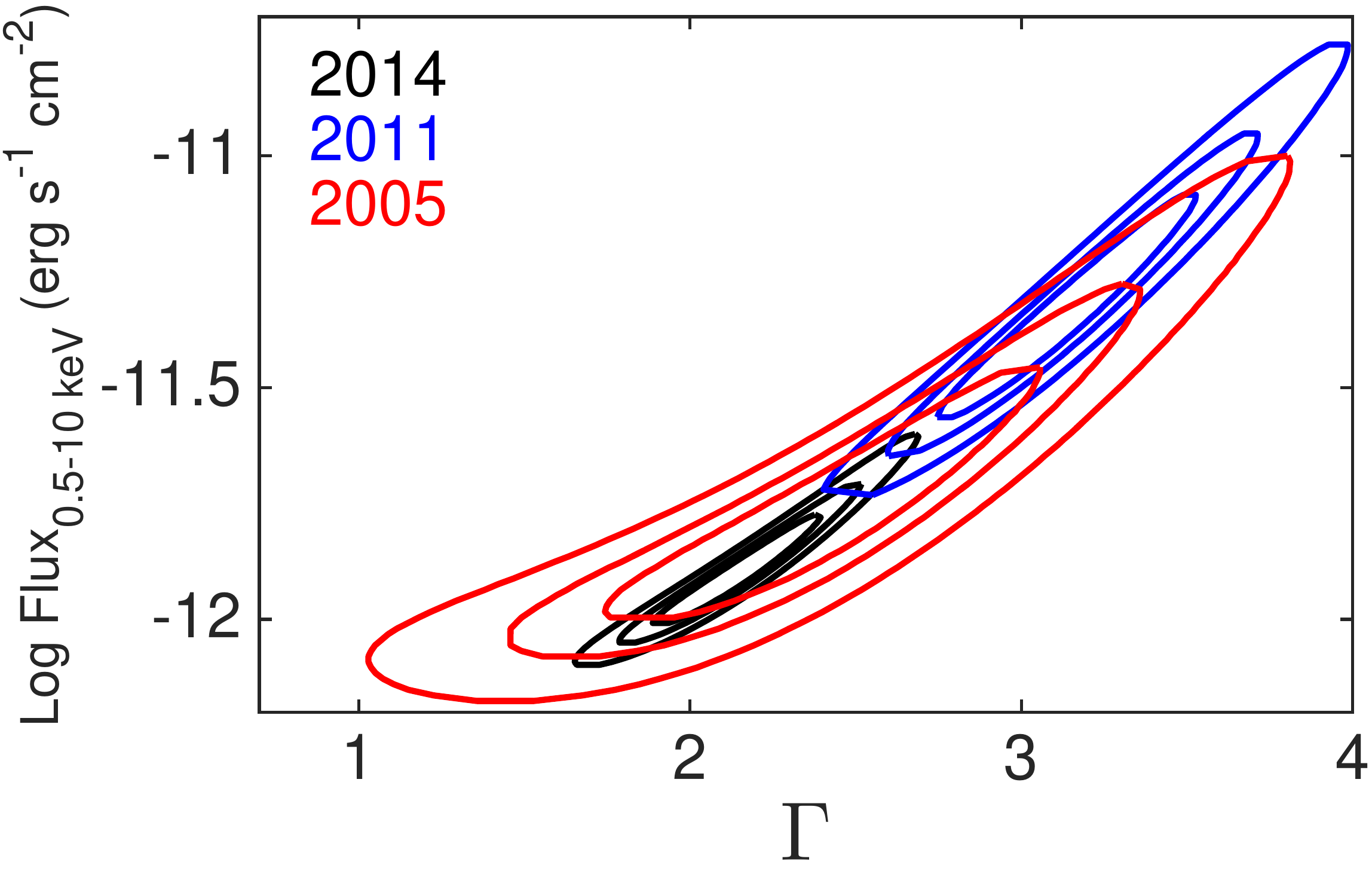}
\caption{$N_{\rm H}$-$\log F_{\rm 0.5-10~keV}$ 1, 2, 3$\sigma$
  contours for the 2014 observations (black lines), 2011 observation 
  (blue lines), and the 2005 observation (red lines). See text for
  details.}
\label{conAll}
\end{center}
\end{figure}

\subsection{Extended emission long-term properties}

The field of \src\ has been observed twice with \xmm\ in the past,
first in September 2005 and later in August 2011, 40 days after the
source went into outburst (Y+12). To understand whether the extended
emission varied between all \xmm\ observations, we fit the 0.5-10~keV
spectra of these older observations simultaneously with the 2014
observations\footnote{For the details on the spectral extraction of
  the 2005 and 2011 spectra, we refer the reader to Y+12. We also note
  that the 2014 observations have significantly higher S/N ratio
  compared to the 2005 and 2011 observations.}. We bin all spectra to
have 5 counts per bin and use the C-stat for spectral fitting. We link
all parameters together except for the PL photon indices and
normalizations. This resulted in a C-stat of 4875.60 for 4677
d.o.f. We find a hydrogen column density $N_{\rm
  H}=(8\pm1)\times10^{22}$~cm$^{-2}$. In Figure~\ref{conAll}, we 
show the $N_{\rm H}$-$\log F_{\rm 0.5-10~keV}$ contours (1, 2, and
3$\sigma$) from all three episodes (2014-black, 2011-blue, 2005-red).
All observations are consistent with one another at the 3$\sigma$
level. Compared to the 2005 and the 2014 observations, the 2011
observation shows, on average, a softer spectrum and a larger
flux. The 2011 observation, however, was 40 days after \src\ went into
outburst. The magnetar spectrum is soft during that observation
($\Gamma=4.2$), and its high flux caused a bright, even softer, dust
scattering halo detected with \xmm\ (Y+12) as well as with \chandra\
\citep{esposito13MNRAS:1834}. Hence, the 2011 extended emission
spectrum is likely contaminated by these two components. We conclude
that the nebula flux and spectral curvature is consistent with being
constant over a span of 9 years from 2005 to 2014.

\subsection{\src}

We derived a rough estimate of the \src\ flux during obs.~1
and a $3\sigma$ upper limit during obs.~2, i.e., about 950 and 1160
days since the August 2011 outburst. Assuming an absorbed BB
model with parameter values similar to the ones derived at late
stages of the outburst (\citet{esposito13MNRAS:1834}; $kT=0.6$~keV and
$N_{\rm H}=12\times10^{22}$~cm$^{-2}$), we find a background-corrected
unabsorbed 0.5-10 keV flux $F_{\rm
  0.5-10~keV}\approx4.0\times10^{-14}$~erg~s$^{-1}$~cm$^{-2}$ and
$F_{\rm 0.5-10~keV}\lesssim3.0\times10^{-14}$~erg~s$^{-1}$~cm$^{-2}$,
for obs.~1 and obs.~2, respectively. This upper-limit is comparable to the
one derived by \citet{kargaltsev12apj:1834} using a 2009 \chandra\
observation.

\section{Discussion}
\label{discuss}

\subsection{Extended emission: scenarios without a wind nebula}

{\bf Dust scattering halo.} Scattering of soft X-ray photons by dust
in the line of sight to magnetars is a common phenomenon due to 
heavy absorption in their direction \citep[e.g.,
][]{tiengo10ApJ:1547}. The hydrogen column density toward \src\ and
its surrounding extended emission is of the order of
$10^{23}$~cm$^{-2}$, enough to cause a dust scattering halo
in the presence of a bright illuminating source. When \src\ went into outburst in September 2011, its flux increased by
more than 3 orders of magnitude compared to its quiescent flux \citep{
  kargaltsev12apj:1834}. That caused the detection of a dust
scattering halo around the magnetar in \chandra\
\citep{esposito13MNRAS:1834} and \xmm\ (Y+12). \chandra\ observations
throughout the outburst indicated that the dust scattering halo
suffered little delay in its flux decay compared to \src. This placed
the magnetar $\sim$200 pc away from the dust cloud causing the halo
\citep{esposito13MNRAS:1834}. The halo detected with \chandra\ had a
size of about 30\arcsec. Emission from dust at larger angular
distances from the source are expected to suffer a delay according to:

\begin{equation}
\theta(t) \approx \left[\frac{2c}{d}~\frac{1-x}{x}~t\right]^{1/2},
\end{equation}

\noindent where $\theta(t)$ is the off-axis angle to the observer at
time $t$, $d$ is the distance from the observer to the source, and
$x=d_{\rm   dust}/d$ where $d_{\rm dust}$ is the distance from the
observer to the dust screen \citep{trumper73AA:dust}. The observed
angle $\theta$ is related to the scattering angle $\theta_{\rm scat}$
through $\theta_{\rm scat}=\theta/(1-x)$, considering that the
scattering angles are usually small enough ($\theta_{\rm
  scat}\lesssim10$\arcmin, \citealt{trumper73AA:dust}). The flux decay
from dust at a given scattering angle follows three branches,
depending on the scattering grain size ($a$) and the energy of the
incident photon ($E$, see equations 8 and 9 of
\citealt{svirski11MNRAS:dust1806}); a constant interval where the
scattering is dominated by the largest grains (e.g.,
$a\sim0.3-1~{\rm \mu m}$) followed by a steep power-law decay for
intermediate size grains and an exponential decay for scattering from
the smallest grains \citep{svirski11MNRAS:dust1806,vasil2015:v404dust}.

Considering that $d=4$~kpc, $d_{\rm dust}=3.8$~kpc \citep[based on
][]{esposito13MNRAS:1834}, and $\theta\approx2$\arcmin, we find a
scattering angle $\theta_{\rm scat}\approx43$\arcmin. The largest
scattering angle that corresponds to the constant flux branch is approximated as
$\tilde{\theta}_{\rm scat,~max}=10.4/[(E/1~{\rm keV})\times(a_{\rm
  max}/0.1~{\rm \mu m})]$~arcminutes \citep{mauche86ApJ:halo}, which,
for a grain size $a=0.3~{\rm \mu m}$, corresponds to 3.5\arcmin. This
indicates that the X-ray emission, if due to scattering from dust, is
beyond the constant flux regime and in the steep PL decay regime, in
contrast to the constant flux we calculate between 2011 and March and
October 2014.

Another way of looking at the problem is considering the time delay
corresponding to the onset of the steep PL decay for a given
scattering angle, i.e., the time over which the flux from the
scattered dust is constant and dominated by scattering from the
largest grains. This is given by \citep{svirski11MNRAS:dust1806},

\begin{equation}
\begin{split}
t(a_{\rm max}) & \approx \frac{dx(1-x)}{2c}~\tilde{\theta}^{2}_{\rm scat}  \\
                    & \leq 3.7\times 10^{3}D_{\rm 4~kpc}
                    \left(\frac{E}{1 {\rm keV}}\right)^{-2}~\left(\frac{a_{\rm max}}{1{\rm \mu m}}\right)^{-2} ~{\rm s},
\end{split}
\end{equation}

\noindent where the inequality uses the fact that $x(1-x)\leq 1/4$. Again, this
is inconsistent with the  constant flux we calculate between 2011 and
March and October 2014.

An additional argument against the dust scattering halo interpretation
is the spectral shape of the extended emission. The cross-section
of the dust grains scales as $E^{-2}$ of the incident photon energies
\citep[e.g., ][and references
therein]{trumper73AA:dust,rivera10ApJ:halo}. Hence, for a source with
a PL spectrum $E^{-\Gamma}$, illuminating a spherical dust
distribution, the resulting halo spectrum scales roughly as
$E^{-(\Gamma+2)}$, assuming that emission from the entire sphere is
observed. The spectrum of \src\ below 10~keV at the time of the
outburst was soft with $\Gamma=3-4$, which should result in an even
softer halo spectrum. Even if we assume that \src, similar to other
magnetars, possessed a hard X-ray tail above $10$~keV with $\Gamma\sim
1$ that was reprocessed by the dust sphere, a halo scattering spectrum
would still be too soft to reconcile with the $\Gamma\approx 1-2$ we
derive for the inner and outer rings of our extended emission.  This
is true both when the rings were to be associated with distinct,
concentric dust spheres of different radii, or with structured portions of
a single scattering sphere.

{\bf Emission from the SNR W41.} The mixed-morphology class of SNRs is
of interest to the discussion of the extended emission seen around
\src. These represent the class of SNRs where the radio emission
shows a shell-like morphology whilst the X-ray emission is
centrally-peaked \citep[see ][for reviews and references therein]{
  rho1998ApJ:SNR,vink12AARv:SNR}. These SNRs are mostly seen in dense
environments, often in the presence of molecular clouds. Interaction
between these SNRs and the molecular clouds manifests through the
presence of OH masers. These properties are qualitatively in agreement
with the environment of the extended emission we see. The extended
X-ray emission is central to the SNR W41. Analysis of CO observations
indicates a considerable amount of molecular material in this direction
and OH masers have been reported indicating the possible interaction
of the cloud with the SNR \citep{frail13ApJ:1834}. However, the X-ray
emission from this class of SNRs is purely thermal with an average
temperature of about 0.6~keV, and their X-ray spectra show strong
emission lines from metal-rich plasma, in contrast with the
featureless, non-thermal spectrum of the extended emission we see
here. Moreover, these SNRs are usually younger than the $10^5$~yr
estimated age of W41 \citep{tian07ApJ:1834}. Finally, while centrally
peaked, the X-ray emission from these SNRs, is generally present
throughout the radio shell albeit at lower surface brightness
\citep{rho1998ApJ:SNR}. The extended emission we observe is detected
at a very small volume compared to the SNR W41 volume, only at the
position and around the magnetar \src. Hence, the X-ray extended
emission we observe here is inconsistent with a mixed-morphology SNR
origin.

{\bf X-ray reflection nebula.} Giant Molecular Clouds (GMCs) in the
Galactic center region emit X-ray radiation that is thought to be the
reflection of past activity from the supermassive black-hole
Sgr~A$^*$ a few hundred years ago \citep[e.g.,
][]{ponti10ApJ:sgra}. The observational properties of the X-ray
emission from these clouds are a hard 2-10~keV spectrum with a photon 
index $\Gamma\approx1.0$, and a broad (equivalent width of
$\sim1$~keV) neutral or low-ionized fluorescent Fe~K emission line
with a flux proportional to the flux of the illuminating source
\citep{sunyaev98MNRAS:sgrA}. Moreover, the X-ray flux from these GMCs
is typically observed to vary on timescales of years \citep[e.g.,
][]{ponti10ApJ:sgra}. This variability is
more pronounced if the energy in the illuminating source is the result
of a brief intense flare compared to a steady output
\citep{sunyaev98MNRAS:sgrA}. Similar results are derived for other
GMCs with different illuminating sources \citep[e.g., 
][]{sekimoto00PASJ, corcoran04ApJ:etacar}. Hence, a past 
strong bursting episode and/or a Giant Flare (GF) from \src\ could in
principle result in an X-ray reflected spectrum from the GMC in its
direction. However, the X-ray photon index of our extended emission,
$\Gamma=2.2$, is much larger than the hard X-ray spectrum  expected
from reflection. Assuming similar properties between the GMC in the
direction of \src\ and the ones in the Galactic center, a strong
bursting episode or outburst from the source should give rise to a
strong Fe~K line, which we do not detect in the current
observation. Moreover, we do not observe any variability in a span of
9 years, while variability on timescales of a few years have been
reported for the GMC in the Galactic center region
\citep{ponti10ApJ:sgra,terrier10ApJ:sgra}. Finally, assuming
reflection from the GMC, it is hard to reconcile the small angular
size of our extended emission with the angular size of the cloud that
is a few times larger \citep{esposito10MNRAS:0418,tian07ApJ:1834}. We
conclude that a reflection scenario is inconsistent with the X-ray
observational properties of the extended emission we detect around
\src.

{\bf Emission from a background galaxy cluster.} In the X-ray band,
galaxy clusters are observed as extended sources with spectra best fit
by optically-thin thermal models. The gas responsible for their X-ray
emission is mostly found to have a temperature of the order of a few
keV and abundances in the range $\sim 0.4-1.0$~solar \citep[e.g., ][]{
  white00MNRAS:gc,maughan08ApJS:gc}. While most of these galaxy
clusters are seen outside the Galactic plane, \citet{townsley11ApJS}
discovered a galaxy cluster in the Galactic plane (Galactic latitude
$l\sim-1.2\deg$). Compared to the galaxy cluster class, the unusually
high temperature ($kT\gtrsim32$~keV, assuming solar abundance) we
derive when fitting the extended emission around the magnetar with a
thermal model is inconsistent with the temperatures of other
clusters. Moreover, allowing the abundance to vary, the very low value
we derive ($<0.1$, $3\sigma$ confidence) is also inconsistent with
most clusters (even when we consider the different abundance values at
different redshifts). We, therefore, exclude the possibility that the
extended X-ray emission around \src\ is due to a background galaxy
cluster.

\subsection{A magnetar wind nebula}

The asymmetric morphology of the extended emission we detect around
\src, its non-thermal origin with an X-ray PL photon index
$\Gamma\approx2$ and with a surface brightness peaking at the \chandra\
position of the central object \citep{kargaltsev12apj:1834}, as well as its flux
constancy over a 9~yr period, closely resemble the properties of a
number of typical PWNe around RPPs \citep[see,
e.g.,][]{kargaltsev08PWN,kargaltsev13:PWN,bamba10ApJ:pwn}.
The central object in our case, however, is a typical magnetar
source. Hence, given that it appears unlikely that this extended
emission is the result of dust scattering, emission from W41,
reflection, or a background galaxy cluster, we conclude that the 2014
{\it XMM-Newton} data combined with the two earlier observations,
unambiguously confirm our earlier results (Y+12) that the extended
emission around \src\  is the first manifestation of a wind nebula
around a typical magnetar.

The timing properties of the magnetar result in a rotational energy
loss $\dot{E}_{\rm rot}=2.1\times10^{34}$~erg~s$^{-1}$. Assuming that
the X-ray nebula is powered by the magnetar rotational energy, this
translates into an unusually high X-ray efficiency $\eta_{\rm
  X}=L_{\rm X,PWN}/\dot{E}_{\rm rot}=0.1$. 

Compared to the PWN around the high-B RPP PSR~J1846-0258, the wind
nebula around \src\ seem markedly different, with the only shared
observational characterestic being the X-ray spectral curvature with a
common photon index $\Gamma=2.0$ \citep{ng08ApJ:psr1846}. The size of
the PWN around PSR~J1846-0258 is a few tens of arcseconds, an order of
magnitude smaller than the size of the wind nebula around \src. Its
X-ray efficiency of about $2\%$, while at the high end of PWN/RPP
systems, is 5 times smaller than the case of \src. The observational
properties of the PWN around PSR~J1846-0258 seem to indicate a typical
rotation powered PWN in a young system (spin down age
$\tau\approx800$~yrs), in contrast to \src.

Wind nebula around other high-B sources have also been suggested.
\citet[][see also \citealt{rea09ApJpwn}]{camero13MNRAS:rrat1819}
discussed the case of the extended emission around
RRAT~J1819$-$1458 ($B\approx5.0\times10^{13}$~G). Its spectrum can be
well modeled by a PL with a photon index $\Gamma\approx3.5$, much
softer than typical PWN and the wind nebula around \src. Assuming a
nebula origin for the extended emission, a high X-ray efficiency of
about $15\%$ is required. Recently, \citet{israel16mnras:1935} hinted
at the possibility of a wind nebula around the newly discovered
magnetar SGR~J1935+2154 during outburst
($B\approx2.2\times10^{14}$~G). The extended emission is well modeled
by a PL with a soft spectrum, $\Gamma\approx3.0$, and its X-ray
efficiency is about $35\%$, both larger than the case of \src. We
note, however, that in the above two cases, the current available data
are insufficient to draw any firm conclusions on the true nature of
these two extended emission and a dust scattering halo interpretation
is still a viable explanation.

\begin{figure}[t]
\begin{center}
\includegraphics[angle=0,width=0.48\textwidth]{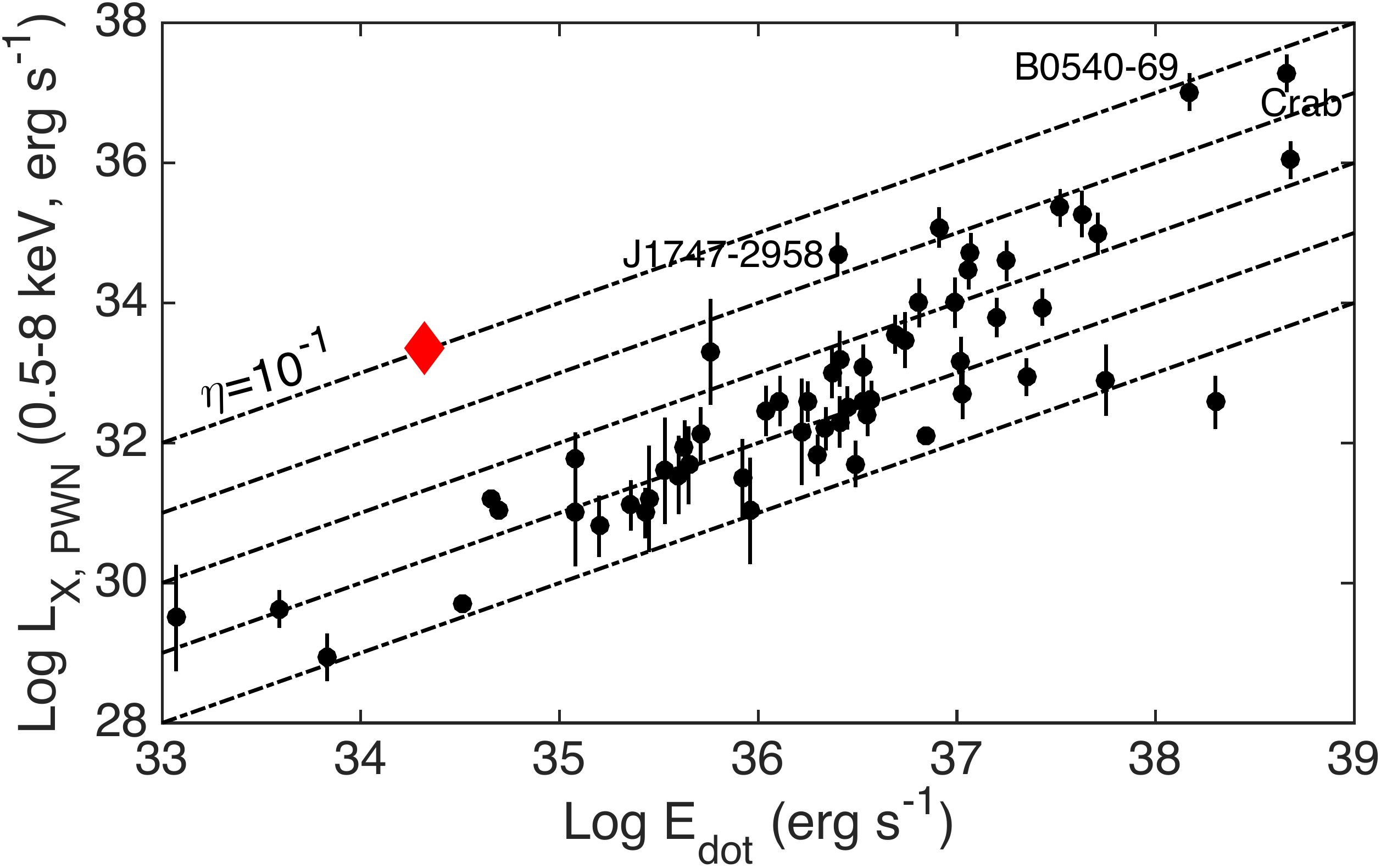}
\caption{PWN X-ray luminosity as a function of the rotational energy
  loss of their powering pulsar. Black dots are RPPs with properties
  taken from \citet{kargaltsev13:PWN}. The dot-dashed lines correspond
  to constant X-ray efficiencies from $10^{-1}$ down to $10^{-5}$ (top
  to bottom) of converting the rotational power to PWN X-ray
  luminosity. The red diamond corresponds to the value derived for the
  wind nebula around the magnetar \src. The names of three other PWN/RPP
  systems are indicated, of which B0540-69 and Crab share the same high
  efficiency as the wind nebula around the magnetar. All data points
  include a statistical as well as a systematic error in their wind
  nebula X-ray luminosity (assumed 40\%). For the magnetar this is
  represented by the size of the diamond. No error is shown on $E_{\rm
  dot}$. Figure adapted from
  \citet{kargaltsev13:PWN}.}
\label{efficiency}
\end{center}
\end{figure}

Figure~\ref{efficiency} shows the X-ray luminosity of PWNe as a
function of the rotational energy loss of their powering pulsar, while
the magnetar is indicated with a red diamond. Only two PWN/RPP systems
come close to the high efficiency of \src: the Crab pulsar, and
B$0540-69$ which is also known as the ``Crab twin''. However, both
sources are much younger with spin-down ages of 1.3 and 1.6~kyr,
respectively, compared to $\tau=4.9$~kyr for the magnetar (the
discrepancy is more pronounced if we assume that \src\ is associated
with the SNR W41 with an estimated age of $\gtrsim50$~kyr). Another
interesting source to mention in this regard is  the PWN/RPP system
PSR~J1747$-$2958 \citep{gaensler04ApJ:mouse}, which has an estimated
age of $\sim$25~kyr and an X-ray efficiency $\eta=0.02$, which are a
factor of 5 larger and lower than the age and efficiency we derive for
\src, respectively. There are  differences between the two systems,
nonetheless. The PWN around PSR~J1747$-$2958 is a prominent nebula in
radio \citep{gaensler04ApJ:mouse}, in contrast to the wind nebula we
see  here \citep{kargaltsev12apj:1834}, while its rotational energy
$\dot{E}$ is more than 2 orders of magnitude larger. In fact,
Figure~\ref{efficiency} shows that all the PWN/RPP systems with
$\dot{E}$ within an order of magnitude of the value for \src\ have
X-ray efficiencies at least two orders of magnitude lower.

Another unusual property of this wind nebula around \src\ is the
extremely high ratio of the nebula X-ray luminosity to the luminosity
of the central magnetar in quiescence, $L_{\rm X,PWN}/L_{\rm
  X,magnetar}\gtrsim40$. \citet[][see also \citealt{kargaltsev08PWN}]{
  kargaltsev07ApJ:1800} found that this ratio is tightly clustered
around 4 for almost all PWN/RPP systems, regardless of efficiency and
age. This value is an order of magnitude lower than the one we derive
here. This high brightness could be the effect of its collision with the W41 SNR reverse
shock (assuming connection between the two). As is shown in
\citet{gelfand09ApJ:PWN}, the compression of the nebula by the reverse
shock tends to increase its particle and magnetic energy, as well as
the strength of its magnetic field. This tends to rapidly increase the
nebula synchrotron luminosity.

An alternative possibility for the unusual high efficiency and
brightness in the wind nebula around \src\ is that there might be an
extra source of power in addition to the rotational energy of the
magnetar. This extra source of power is most likely the decay of the magnetar's ultra-strong magnetic field, with an
inferred surface dipole magnetic field of $B=2.1\times10^{14}$~G.
Particle outflows, either steady or released during bursting episodes,
could be driven out from the magnetar as Alfv\'en waves
\citep{thompson98PhRvD:mag,harding99ApJ:mag,tong13ApJ:wind}. The best
observational evidence are the transient radio nebulae detected from
the magnetars SGR~1900$+$14 and SGR~1806$-$20 following their
respective 1998 and 2004 GFs \citep{frail99Natur:sgr1900,
  gaensler05Natur:1806,gelfand05ApJ:sgr1806,taylor05ApJ:1806,
  granot06ApJ:1806}. It is obvious that the nebula we see around \src\
is steady over a minimum of a 9~yr period, hence, intrinsically
different from the radio nebulae seen around SGR~1900$+$14 and
SGR~1806$-$20 following the GFs. Particle outflows in magnetars,
however, are not restrained to GF emission, and are expected during
regular bursting episodes  \citep{thompson96ApJ:magnetar,
  gill10MNRAS:SGRGF}, particularly given that the magnetic Eddington
limit is low enough that it can be breached even by the short bursts
\citep{watts10ApJ:PRE,vanputten13MNRAS:magAtmos}. Indeed, the strong
torque changes seen in many magnetars point towards a particle wind
escaping out through open field lines out to (at least) the light
cylinder \citep{kaspi14ApJ:1745,archibald15ApJ:1048,
  younes15ApJ:1806}. Moreover, the discovery of quasi-periodic
oscillations in the short-recurring bursts of magnetars point to
Alfv\'{e}n waves driven by ongoing seismic activity
\citep{hupponkothen14ApJ:rxte,hupponkothen14ApJ:fermi}. \citet{harding96AIPC}
studied the case of particle outflow following short magnetar bursts
and found that the cooling time-scale is very long (compared to the
time between bursts) such that the nebular emission is steady rather
than transient, in agreement with our results.

The above qualitative theoretical reasoning raises the question about why would
\src\ be the only magnetar so far powering a wind nebula, given that previous
searches around individual magnetars have returned no sign of extended
emission attributable to wind nebulae \citep[e.g., ][]{vigano14:1806}.
With only one observed so far, it is difficult to draw any firm
conclusions. Nevertheless, \src\ has some interesting characteristics
that are not shared with the entire magnetar population. First, the
environment of \src\ is extremely crowded, with a \fermi\ GeV, a
\hess\ TeV source, a SNR, a GMC, and an OH maser in its vicinity
\citep{frail13ApJ:1834,hess15AA:1834}. The relationship between all
these sources is unclear. However, it is tempting to speculate that
environmental effects from such a rich field could be playing a role
in the production of this wind nebula (e.g., triggering of pair
cascade by external gamma-rays from a nearby source; \citealt{
  shukre82ApJ,istomin11JETP}). Second, the \src\ X-ray 
luminosity in quiescence is $L_{\rm X}=5\times10^{31}~D^2_{\rm
  4~kpc}$~erg~s$^{-1}$. Only five other magnetars (SGR~0418$+$5729,
SGR~1745$-$2900, XTE~J1810$-$197, Swift~J1822.3$-$1606,
3XMM~J185246.6$+$003317\footnote{http://www.physics.mcgill.ca/~pulsar/magnetar/main.html}) 
have luminosities $\lesssim10^{32}$~erg~s$^{-1}$. Among these 5, three
have the smallest surface $B$ fields measured (SGR~0418$+$5729,
Swift~J1822.3$-$1606, 3XMM~J185246.6$+$003317; $B<4\times10^{13}$~G),
and only one source, SGR~1745$-$2900, has a rotational energy loss
rate $\dot{E}$ similar to \src, while the rest have $\dot{E}$ at least
an order of magnitude lower. Hence, from an observational point of
view, it seems that the combination of very weak X-ray luminosity, a
magnetar-like B-field strength, and a somewhat large $\dot{E}$
(properties that are only shared by the Galactic center magnetar
SGR~1745$-$2900) may favor of a wind nebula
production. Another possibility is that the \src\ magnetar/nebula
system is an older analogue to the Kes~75 system, where the
central pulsar evolves into a magnetar while preserving its originial
pulsar wind nebula (PWN).

Finally, we briefly discuss the softening trend within the \src\
nebula (see Section~\ref{specana}). While the usual cause for spectral
differences within wind nebula around pulsars is a change in spectral
curvature rather than absorption, this latter scenario should be
considered in our case. The heavy absorption for \src\ is thought to
be due to the existence of a GMC in its direction
\citep{tian07ApJ:1834}. The densities within these clouds could be
non-uniform causing spatially variable absorption. The magnetar,
surrounded by the inner nebula, was a bright X-ray point source in
2011 when in outburst. The \xmm\ spectrum was fit in Y+12 with an
absorbed PL ($N_{\rm H}=(24\pm1)\times10^{22}$~cm$^{-2}$) and an
absorbed black-body (BB, $N_{\rm
  H}=(13\pm1)\times10^{22}$~cm$^{-2}$). Y+12 found that the absorbed
BB fit ($\chi_\nu^2=1.04$ for 232 d.o.f.) was superior to the PL fit
($\chi_\nu^2=1.15$ for 232 d.o.f.), possibly indicating that the true
absorption towards the magnetar, and by extrapolation the inner
nebula, is similar to the absorption towards the outer nebula. This
potentially points to a change in the photon index as the primary
cause for the spectral softening that we see in our data,
strengthening the case for a typical wind nebula around \src, similar
to PWNe around young pulsars. Such softening trend could be attributed
to synchrotron burnoff of energetic particles. This conclusion,
however, requires a deeper X-ray observation to firmly confirm it
and/or a theoretically-motivated spectral modeling (i.e., whether
emission from marginally fast cooling electrons with a given initial
power-law energy distribution could quantitatively explain the spectral
softening observed).

\section{Conclusion}

In this paper, we studied two deep \xmm\ observations of the extended
emission around the magnetar \src. The observations, separated by 7
months, were taken in March 2014 and October 2014, 2.5 and 3.1 years
after the source went into outburst. The magnetar is weakly detected
in the first observation, while it faded below detection limit
during the second one. The extended emission is clearly detected in
both observations; it is best described with a non-thermal PL model
with a photon index $\Gamma\approx2.2\pm0.2$. Fitting these spectra
with archival ones taken 3 and 9 years earlier, we find that the flux
and spectral curvature of the xtended emission is constant with
$F_{\rm 0.5-10~keV}\approx1.1\times10^{-12}$~erg~cm$^{-2}$~s$^{-1}$.
This provides a strong observational case confirming the Y+12 results
that \src\ is indeed the first magnetar to show a surrounding wind
nebula.  Our results imply that these properties are likely no longer
exclusive to RPPs, and further narrow the gap between these two
sub-populations of isolated NSs. A more in-depth theoretical
interpretation of these results will be presented in a separate
accompanying paper.

\section*{Acknowledgments}

Based on observations obtained with {\it XMM-Newton}, an ESA science mission
with instruments and contributions directly funded by ESA Member
States and NASA. ALW acknowledges support from NWO Vidi Grant
No. 639.042.916. O.J.R. acknowledges support from Science Foundation
Ireland under Grant No. 12/IP/1288. CK and GY acknowledge support by
NASA through grant NNH07ZDA001-GLAST (PI: C. Kouveliotou). GY thanks
Dale Frail for insightful discussions on this project. We thank the
referee for useful comments that improved the quality of the
manuscript.

\end{document}